# THE IMPACT OF SMEs ON EMPLOYMENT CREATION IN MAKURDI METROPOLIS OF BENUE STATE

[1]Bridget Ngodoo Mile, [2]Victor Ushahemba Ijirshar and [3]Mlumun Queen Ijirshar

[1,2,3]Benue State University, Makurdi-Nigeria.

[1]milengodoo@gmail.com, [2]ijirsharvictor@gmail.com [3]queenmlumun@gmail.com

**Abstract**

*SMEs remain a veritable tool that generates employment opportunities. This study examined the impact of SMEs on employment creation in the Makurdi metropolis of Benue state. A sample size of 340 entrepreneurs was chosen from the population of entrepreneurs (SMEs) in the Makurdi metropolis. The study used logistic regression to analyse the impact of SME activities on employment creation or generation in the state and found that SMEs contribute significantly to employment creation in the state but are often faced with the challenges of lack of capital, absence of business planning, lack of confidence in the face of competition, unfavorable environment for the development of SMEs, high government taxes and inadequate technical knowledge. The study therefore recommended that the government should implement capital or credit enhancing programmes and an enabling environment for smooth running of the SMEs. Tax incentives should also be granted to infant enterprises, and tax administration should be monitored to avoid excessive tax rates imposed by tax collectors.*

**Keywords:** Employment creation, Entrepreneurship, Unemployment, and Small and Medium Scale Enterprises (SMEs).

**JEL Classification**: E24, J64, M13

## Introduction

Small and medium scale enterprises (SMEs) play an important role in the economic growth of both developed and developing nations. One of the major goals of any economy is the achievement of full employment, and the attainment of this macroeconomic objective has remained an issue that continues to receive attention in developing countries, particularly those in Africa where high-level poverty exists with an increasing unemployment rate (Oni, 2006). The dynamic role of small- and medium-scale enterprises as engines through which the growth and development objectives of nations can be achieved has been recognized and stated in several studies. According to Kpelai (2009), small-scale enterprises are generally regarded as the driving force of economic growth, job creation and poverty reduction, as they provide the necessary groundwork for entrepreneurship development, innovation and risk-taking behaviour and provide the foundation for long-term growth dynamics and the transition towards larger enterprises. The sector serves as a catalyst for employment generation, national growth, poverty reduction and economic development and can boast of being the major employers of labour compared to the major industries, including multinationals (Kadiri, 2012). Nnanna (2001) also supports that SMEs help in the achievement of improvement in rural infrastructure and improved living standards of rural dwellers, thereby creating employment utilization of indigenous technology, production of intermediate technology and an increase in the revenue base of private individuals and the government (Wahab and Ijaiya, 2006).

Most countries are characterized by a large number of micro and small/medium businesses mainly in the informal sector. Nigeria falls within this category of economies, as micro, small- and medium-sized enterprises account for over 95% of nonoil-producing activities outside agriculture, with small-scale enterprises accounting for approximately 85% of all firms operating in the economy (Nnanna, 2001). The contributions claimed for small-scale enterprises are numerous. According to Safiriyu and Njogo (2012), the vibrancy of small- and medium-scale business subsectors is an important part of an economy because they bring about new ideas and provide job opportunities, thus encouraging entrepreneurship. They have an immediate impact on employment generation (Ayozie and Latinwo, 2010). Thus, the development of SMEs is seen as an accelerator to the achievement of wider economic and socioeconomic objectives, including poverty alleviation and unemployment reduction (Uchechukwu, 2003).

In an attempt to address the unemployment problem in Nigeria, a plethora of strategies and measures have been proposed, including National Development Plans (NDPs) and the rolling plans, the National Directorate of Employment (NDE), the Better Life Programme (BLP) for rural women, the Family Support Programme (FSP) and the Family Economic Advancement Programme (FEAP) during the post-SAP period, the Poverty Alleviation Programme (PAP), which was later changed to the National Poverty Eradication Programme (NAPEP), the Small and Medium Enterprises Equity Investment Scheme (SMEEIS), the Small and Medium Enterprises Development Agency of Nigeria (SMEDAN), and the Bank of Industry (BOI), among others.

Historically, from 2006 until 2011, Nigeria unemployment averaged 14.6%, reaching an all-time high of 23.9% in 2011 from a record low of 5.33% in 2006 (NBS/CBN Surveys, 2014). In Benue state, the unemployment rate is also on the increase evidenced by an average of 14.2% in 2011 (General Household Survey Report/NBS/CBN Surveys, 2014). The 2012 National Bureau of Statistics (NBS) poverty profile depicts Benue State with an unemployment rate of 52.4%. This has posed a great threat to the economic growth and employment status of Benue State.

However, the government programmes and initiatives led to the proliferation of SMEs in Benue State and Makurdi Metropolis in particular. Thus, with an increased number of SMEs, it was expected that these small- and medium-scale enterprises would expand and metamorphose into large enterprises with large capital, productivity, profit and enhanced job opportunities. The phenomenon of a high unemployment rate despite the increased numbers of SMEs raises questions yet to be answered bearing in mind the prominence theories accorded to the role of SMEs in employment creation. Against this background, this study seeks to investigate the impact of SMEs on employment creation in the Makurdi metropolis. The study therefore seeks to answer the following questions:

i. What are the socioeconomic characteristics of SME operators in the Makurdi metropolis?
ii. To what extent have SMEs contributed to employment creation in the Makurdi metropolis?
iii. What are the factors limiting the effective performance of small-scale enterprises in the Makurdi metropolis?

This study is a point analysis, as it is restricted to the contribution of SMEs to employment creation, with particular interest in the Makurdi metropolis.

**Conceptual Issues**

*Small and Medium Scale Enterprises***:**

In Nigeria, the Central Bank of Nigeria (CBN), in its monetary policy circular No 22 of 1988, defined small-scale enterprises as having an annual turnover not exceeding five hundred thousand naira (Ali 2003), as cited in Ogbuabor, Malaolu and Elias (2013). The Federal Ministry of Industries (1973) defined small-scale enterprises as businesses that have total capital (land, building machinery equipment and working capital) of up to N60,000 and employ up to 50 persons. Ifechukwu (2000) views a small-scale enterprise as a business operated mainly with hired labour usually not exceeding 50 workers if no motive power is used. Generally, such businesses are characterized by a labour intensive mode of production, flexible operation as they adjust quickly to various factors, use indigenous raw materials, localized operations, a low gestation period, and a low level of education/skills (Ifechukwu, 2000; Ali 2003). The Ministry of International Trade and Industry (2005) cited in Jamil and Mohamed (2011) defined SMEs in the manufacturing sector to include companies with larger operating capacity, firms with less than 50 full-time employees and annual turnover of not more than RM10 million (Akinruwa, Awolusi and Ibojo, 2013). In the same vein, Jasra, AsifKhan, Hunjra, Rehman and Zam (2011) defined SMEs as those enterprises that employ no more than 250 employees and work on a small scale. He goes further to show that the technical definition varies from country to country but is usually based on employment, assets, or a combination of the two.

*Employment Creation*

It is very difficult to find a clear definition of employment creation. In every economy, there are both active and inactive populations. The economically active populations are referred to as the employed population. They include those actively engaged in the production of goods and services. Generally, employment is a situation where a person is doing a job to earn income. Employment can also exist where a person is self-employed. Therefore, employment creation can be referred to as the provision of job opportunities for those who are willing and able to work.

*Entrepreneurship*

According to Tyson, Petrin, Rogers (1994), entrepreneurship is viewed as a person who either creates a new combination of production factors such as new sources of supply and new organizational forms or as a person who is willing to take risks or a person by exploiting market opportunities eliminates disequilibrium between aggregate supply and aggregate demand or as one who owns and operates a business. It is a force that mobilizes other resources to meet unmet market demand, the ability to create and build something from practically nothing, and the process of creating value by putting together a unique package of resources to exploit an opportunity. In Nigeria, many definitions have been affirmed by many authors. Odusina (1975) defines entrepreneurship as the process of using available capital in any form from business endeavours in an open and free market economy for the sole purpose of making profit, and it includes all enterprises in new fields or in older ones at all risk levels. Onyido (2011) also opines that entrepreneurship is the practice of starting new organizations or revitalizing nature organizations, particularly new businesses, generally in response to identified opportunities. Entrepreneurship according to this study is seen as the willingness and the ability of an individual to seek out an investment opportunity, establish an enterprise based on this and run it successfully.

*Unemployment*

The International Labour Organization (ILO) defined the unemployed as members of economically active populations who are without work but available for work, which naturally results in the displacement of labour and finally causes unemployment. Oladele et al. (2011) found that, regardless of the type and cause of unemployment, entrepreneurship development is the answer.

**Theoretical Literature**

Resource constraints in developing countries necessitate prioritization as to where to invest first. Unbalanced growth theory specifies that the key sectors for initial investment should be determined on the basis of industrial backwards and forward linkages (Hirschman, 1958). SME activities have significant backwards and forward linkages to enhance income generation and employment creation and are capable of breaking the vicious cycle of poverty in the study area. Therefore, this study is anchored on the theories of critical minimum effort and the unbalanced growth model. The critical minimum effort presupposes that poverty is a serious human problem that is self-perpetuating. Thus, efforts must be geared towards supporting active participation in entrepreneurial activities if poverty must be reduced. SMEs can be supported through the provision of finance to a level at which sustained development could be maintained to achieve that transition from a small scale to a larger one where one can expect steady secular growth and employment creation.

**Empirical Review**

Ogbuabor, Malaolu and Elias (2013) examined the efficacy of tackling the twin economic challenges of poverty and unemployment in Nigeria through the small-scale enterprise commonly known as burnt bricklaying. The study also examined the socioeconomic characteristics of the bricklayers as well as the major challenges militating against their growth and performance. The results indicated that burnt bricklaying has a significant positive impact on poverty alleviation, job creation, and income generation in Nigeria. The study therefore recommended that the challenges of poor infrastructure, low prices of bricks, low demand for bricks, and low operating capital faced by these small-scale enterprises should be addressed by the various tiers of government and the financial system as a viable means of job creation, poverty alleviation and income generation in Nigeria.

Kadiri (2012) examined the contributions of small- and medium-scale enterprises (SMEs) to employment generation in Nigeria. The study provided a sectoral analysis of the efficacy of SMEs as a vibrant tool for employment generation in the country. Binomial logistic regression analysis was employed as a tool for statistical analysis. The study showed that the sector was unable to achieve its goal due to its inability to obtain adequate business finance for the sector. It was also observed that virtually all the SMEs that were sampled relied on informal sources of finance to start their business. The study therefore suggested the need for the integration of the formal activities with those of informal financial institutions. Additionally, the government should, as a matter of urgency, provide the needed infrastructure, such as roads, water, electricity and the needed enabling environment, as these efforts will reduce the cost of doing business and increase the retained earnings of SMEs, their average monthly income and poverty in the long run.

Safiriyu and Njogo (2012) examined the impact of small- and medium-scale enterprises on the generation of employment in Lagos state using a sample size of 150, out of which one hundred and twenty (120) copies were properly completed and retrieved while thirty (30) copies were not retrieved. The instruments used to gather information for the study included questionnaires and interviews. Two different statistical methods were employed to analyse data for the study; the tools were simple percentage and chi-square ($X2$). The results show that small- and medium-scale enterprises and sustainable development of the Nigerian economy are related, just as promotion of SMEs and improvement in employment generation are related. The study therefore concluded that for a nation irrespective of its economic ideology to achieve meaningful and sustainable development, adequate attention must be given to the widespread of economic activities through entrepreneurship and small- and medium-scale enterprise generation.

**Research Methodology**

The target population of this study is estimated to be over 2400 registered SMEs (Corporate Affairs Commission, 2014). The study adopted a random sampling technique. A sample size of 340 respondents was determined by applying the Yamane (1967) formula. Both descriptive and inferential statistics were used for data analysis. In addition, multivariate logit regression analysis was also used.

*Model Specification*

Adopting the model used by Kadiri (2012) in the study of Small and Medium Scale Enterprises and Employment Generation in Nigeria. The model for the study is stochastically stated below as:

Logit (E) = $\beta_0 + \beta_1 COST + \beta_2 GEND + \beta_3 SIZE + \beta_4 EDUC + \beta_5 INCM + \beta_6 ACCF + \beta_7 LNGT + U_t$.

Where;

$B_0$ = Constant term, $\beta_1 – B_7$ Regression Coefficients, COST = Annual cost to the enterprise, GEND = Gender sensitivity of the owner/enterprise (1 if the enterprise is gender sensitive, 0 if otherwise), SIZE = Household size of the entrepreneur, EDUC = Education of the entrepreneur (1 if the owner attains secondary education and above, 0 if otherwise), INCM = Average annual income accruing to business entrepreneurs, ACCF = Access to Finance (1 if the entrepreneur have access to finance, 0 if otherwise) and LNGT = Length of Years in SME business.

The endogenous variable is a dichotomous or dummy variable with (1) if the SME employs at least one additional person in each year (i.e., the marginal employment level per year is at least one person) and (0) if the SME does not employ at least one additional person in each year (i.e., the marginal employment level per year is zero). Thus, binary logistic regression was used for this study. In this study, β4, β5, β6, and β7 were expected to have positive signs, implying a positive relationship with the dependent variable E. On the other hand, β1, β2 and β3 were expected to be negative, implying an inverse relationship.

*Decision Rule*

The Likelihood Ratio (LR) statistic was used to test the null hypothesis that all the slope coefficients are simultaneously equal to zero (i.e. $\beta_1= \beta_2= \beta_3= \beta_4= \beta_5= \beta_6= \beta_7 = 0$). If LR statistic value is greater than its probability (P) value, we rejected the null hypothesis and accept the alternative that all the β's are not equal to zero. Also, any McFadden R-squared level greater than 0.50 (50%) suggests a strong relationship between the dependent variable (E) and the predictor variables (the X's). Any probability value of the coefficient less than or equal to = 0.05 also implies that the variable is statistically significant at 5% critical level.

**Data Presentation and Analysis**

*Socio – Economic Characteristics of the Respondents*

The categorical statistics of the explanatory variables and dependent variable captured in the model are estimated and analysed in Table 1 and 2 below:

**Table 1: The Categorical Descriptive Statistics of the Explanatory Variables**

|  |  | Mean |  |
| --- | --- | --- | --- |
| Variable | Dep=0 | Dep=1 | All |
| COST | 101960.6 | 105451.3 | 104342.5 |
| GEND | 0.990741 | 0.116379 | 0.394118 |
| SIZE | 9.342593 | 9.314655 | 9.323529 |
| EDUC | 0.212963 | 0.883621 | 0.670588 |
| INCM | 112943.3 | 425091.3 | 289467.8 |
| ACCF | 0.231481 | 0.875000 | 0.670588 |
| LNGT | 5.925926 | 8.280172 | 7.532353 |
| C | 1.000000 | 1.000000 | 1.000000 |
| Observations | 108 | 232 | 340 |

Source: E-views Output

From Table 1 above, the annual average cost of the sampled respondents who were able to employ at least one person per year was ₦105451.30k with ₦425091.30k annual income average while ₦101 960.60k was the average cost of the enterprise that could not offer at least one person additional employment in the study area with the average annual income of ₦112943.30k. This implies that there was a high average annual cost incurred by those who could not employ additional labour each year relative to their average annual income. However, enterprises that were creating additional employment each year had high average annual income with relatively lower average annual cost (comparative ratio analysis). The average annual cost of the sampled enterprises was ₦104342.50k, with an average annual income of ₦289467.80k. This means that the income of an enterprise matters a lot to the successful creation/generation of employment in the study area. The average years of experience of the enterprises that could generate additional employment were approximately 8.2 years, with their holders having a lower average family size of 9.31, while the enterprises that could not offer additional employment in each year had fewer years in operation (5.9), with entrepreneurs having a higher family size of 9.34. The table reveals averages of 7.5 years and 9.3 persons for the years of experience and the family size of the sampled respondents, respectively. The table also reveals that most of the enterprises that had at least one additional employment in each year had access to finance and were more educated on average with less sensitivity in gender (in terms of employment). The dummy variable of gender sensitivity indicates that a greater percentage of 99% with a standard deviation of 0.096 agreed that being gender sensitive may reduce the chances of employment. The dummy variable of education also indicates that 88% of the respondents (i.e., average dummies of 0.88 for Dep=1) with a standard deviation of 0.321 are educated, thus influencing the expansion technique and generating more employment opportunities every year.

**Table 2: The Categorical Descriptive Statistics of Dependent Variable;**

| Dep. Value | Count | Percent | Cumulative Count | Percent |
|---|---|---|---|---|
| 0 | 108 | 31.76 | 108 | 31.76 |
| 1 | 232 | 68.24 | 340 | 100.00 |

Source: E-views Results, 2015

The dummy variable of employment creation in Table 2 indicates that approximately 68% of the SMEs in the study area were able to generate at least one additional employment against a smaller percentage of 32%. This implies that more than 232 new employment opportunities are offered each year by SMEs in the Makurdi Metropolis.

*The Impact of SMEs on Employment Creation in Makurdi Metropolis*

To effectively capture the extent of the change in the dependent variable (employment or job creation) due to the activities of small and medium scale enterprise, the study used logistic regression and the results are presented in Table 3 below:

**Table 3: Logistic Regression results of the model**

| Dependent Variable: E | | | | |
|---|---|---|---|---|
| Variable | Coefficient | Std. Error | z-Statistic | Prob. |
| COST | -0.024449 | 0.006552 | -3.731314 | 0.0002 |
| GEND | -0.495553 | 0.031901 | -15.53425 | 0.0000 |
| SIZE | -0.000417 | 0.001981 | -0.210349 | 0.8335 |
| EDUC | 0.256033 | 0.033179 | 7.716740 | 0.0000 |
| INCM | 1.701851 | 0.617890 | 2.754295 | 0.0059 |
| ACCF | 0.221522 | 0.029300 | 7.560415 | 0.0000 |
| LNGT | 0.824642 | 0.243620 | 3.384950 | 0.0009 |
| C | -14.71734 | 5.596244 | -2.629860 | 0.0085 |
| McFadden R-squared | 0.921527 | Mean dependent var | | 0.682353 |
| S.D. dependent var | 0.466248 | S.E. of regression | | 0.116886 |
| Akaike info criterion | 0.145163 | Sum squared resid | | 4.535935 |
| Schwarz criterion | 0.235256 | Log likelihood | | -16.67777 |
| Hannan-Quinn criter. | 0.181062 | Deviance | | 33.35554 |
| Restr. Deviance | 425.0565 | Restr. log likelihood | | -212.5283 |
| LR statistic | 391.7010 | Avg. log likelihood | | -0.049052 |
| Prob(LR statistic) | 0.000000 | | | |
| Obs with Dep=0 | 108 | Total obs | | 340 |
| Obs with Dep=1 | 232 | | | |

Source: E-views Output

The result from Table 3 indicates that the coefficient of cost incurred in production by the sampled entrepreneurs (COST) is negative (-0.024), which is correctly signed and statistically significant at the 5% level of significance. This implies that the high cost incurred in production by entrepreneurs will reduce the probability of creating employment opportunities in the study area and vice versa. The estimate for gender sensitivity (that is, an entrepreneur being gender sensitive in terms of offering employment) is negative (-0.50), which is correctly signed and statistically significant at the 5% critical level. Gender sensitivity in terms of offering employment tends to decrease the probability of SMEs creating jobs or generating employment in the study area. This is because even when there are qualified men workers and an entrepreneur need women (that is, being gender sensitive), unemployment is created if the women are not found. Therefore, gender sensitivity by entrepreneurs reduces the level of employment creation in the study area.

The coefficient of household size (SIZE) of the sampled respondents is negative (-0.00042), which is correctly signed but statistically insignificant at the 5% critical level. It therefore indicates that the household size of the entrepreneurs in the study area is significant in neither reducing the level of employment creation nor otherwise. However, the level of education (EDUC) of the sampled respondents had a positive (0.26) relationship with job creation and is statistically significant at the 5% critical level. This implies that the level of education of the sampled entrepreneurs had a strong and significant positive influence on the probability of SMEs creating more jobs or generating employment in the study area.

Moreover, the coefficients of income (INCM) and access to finance (ACCF) by the entrepreneurs were positive (1.7 and 0.22, respectively) and correctly signed and statistically significant at the 5% critical level. This implies that higher income obtained by entrepreneurs and greater access to finance by them has a positive influence on the probability of SMEs creating or generating more jobs in the study area.

The length of the business in terms of years of operation (LNGT) was reported to be positive (0.82) and statistically significant at the 5% critical level. This means that the length of business operations has much influence on job creation in the study area.

All the standard errors of the individual variables are minimum except the household size the sampled entrepreneurs (SIZE), thereby producing high Z–statistic and below 0.05 probability values, which indicate that all the variables are statistically significant at the 5% critical level except the household size. The McFadden R2 of 0.92 implied that all explanatory variables included in the model explained total variations in the dependent variable (employment status) by 92%, which indicates a goodness of fit. The LR statistic of 391.7 coupled with a probability (LR Statistic) of 0.00 indicated the reliability of the explanatory variables with regard to the dependent variable, and the minimum value of the standard errors of regression proved the robustness of the model. Moreover, the results of goodness of fit evaluation for binary specification using Andrews and Hosmer–Lemeshow tests indicate 23.174 and 80.3422 for the H–L statistic and Andrews statistic, respectively. These have fully explained the goodness of fit for the logit or binary specification of the estimated model.

Given the decision rule that rejects the null hypothesis if the probability of the LR statistic is less than the critical value, the probability (LR statistic) of 0.0000 is less than the 0.05 critical value; we therefore reject the null hypothesis in favour of the alternative and conclude that SMEs have a significant impact on employment creation in the Makurdi metropolis of Benue State.

*The Factors Militating Against the Effective Performance of SMEs in Makurdi Metropolis*

The following factors militate against the effective performance of SMEs in the study area. The factors are arranged in Table 4 in hierarchical order.

**Table 4 Factors Militating against the Performance of SMEs**

| Factors Militating against the Performance of SMEs | **Percentage(No of Respondents)** |
|---|---|
| 1. Insufficient Capital | 100% (340) |
| 2. Absence of business planning | 94.1% (320) |
| 3. Lack of confidence in the face of competition | 91.8% (312) |
| 4. Unfavorable environment for the development of SMEs (enabling environment) | 89.1% (303) |
| 5. Government Taxes | 78.8% (268) |
| 6. Inadequate technical knowledge | 72.4% (246) |

**Source: Authors' Computation**

Table 4 shows six factors limiting the performance of SMEs in the Makurdi metropolis in their hierarchical order. The last column shows the proportion of the respondents who mentioned the constraints/militating factors. The results show that a lack of capital or inadequate capital to buy stock and equipment was the major challenge for all entrepreneurs. Among other problems were absence of business planning, lack of confidence in the face of competition, unfavorable environment for the development of SMEs, high government taxes and inadequate technical knowledge.

**Conclusion and Recommendations**

This study found that SMEs have a positive impact on generating employment in the study area. The empirical literature reviewed indicated that SMEs remain an important tool that can guarantee employment in the state and Nigeria at large. Thus, SMEs are reliable means and catalysts for job creation, enhanced household income, wealth creation and poverty reduction. The study therefore concludes that SMEs contribute to employment creation for the teaming youths of the state and thus recommends the following:

Since the sources of SMEs' business capital are grossly inadequate, there is an urgent need for the government to implement capital or credit enhancing programmes to broaden the credit base of their institutions, thereby enhancing credit availability at an affordable interest rate. This is because access to finance contributes significantly to employment creation in the area.

Benue state and Nigeria at large should create an enabling environment for the smooth running of SMEs. This is done through establishing infrastructural facilities that would enable effective running of these SMEs, thus creating more employment opportunities in the state and Nigeria at large. These may also include the provision of good roads, electricity, water supply, research institutes and so on. These no doubt will go a long way in reducing the production costs of SMEs.

Tax incentives should be granted to infant enterprises, and tax administration in the state should be monitored to avoid excessive tax rates imposed by tax collectors. This would help encourage investment in the state, thereby creating employment opportunities.